\documentclass[aps,twocolumn,a4paper,floatfix,superscriptaddress,prapplied,longbibliography]{revtex4-1}
\usepackage[toc,page]{appendix}
\usepackage{braket}
\usepackage{epsfig,graphicx,times}
\usepackage{amstext}
\usepackage{amsmath}
\usepackage{amssymb}
\usepackage{mathrsfs}
\usepackage{dcolumn}
\usepackage{bm}
\usepackage[tight]{subfigure}
\usepackage[colorlinks,linkcolor=blue,anchorcolor=blue,citecolor=blue,urlcolor=blue]{hyperref}
\usepackage{color}
\usepackage{siunitx}
\usepackage{appendix}
\usepackage{placeins}
\usepackage{textcomp}
\usepackage{bbold}
\usepackage{float}
\usepackage{verbatim}
\usepackage{minitoc}
\usepackage[dvipsnames]{xcolor}

\usepackage{soul}
\usepackage[framemethod=tikz]{mdframed}

\begin{document}

\title{Phase-controlled robust tripartite quantum entanglement in cavity-magnon optomechanics}

\author{Jing-Xue Liu}
\affiliation{School of Physics, Henan Normal University, Xinxiang 453007, China}
\affiliation{Key Laboratory of Low-Dimensional Quantum Structures and Quantum Control of Ministry of Education, Department of Physics and Synergetic Innovation Center for Quantum Effects and Applications, \\
Hunan Normal University, Changsha 410081, China}
	
\author{Ya-Feng Jiao}
\email{yfjiao91@foxmail.com}
\affiliation{School of Electronics and Information, Zhengzhou University of Light Industry, Zhengzhou 450001, China}
\affiliation{Academy for Quantum Science and Technology, Zhengzhou University of Light Industry, Zhengzhou 450002, China}

\author{Bin Yin}
\affiliation{Key Laboratory of Low-Dimensional Quantum Structures and Quantum Control of Ministry of Education, Department of Physics and Synergetic Innovation Center for Quantum Effects and Applications, \\
	Hunan Normal University, Changsha 410081, China}

\author{Hong-Yun Yu}
\affiliation{College of Advanced Interdisciplinary Studies, NUDT, Changsha 410073, China}

\author{Ruo-Chen Wang}
\affiliation{School of Electronics and Information, Zhengzhou University of Light Industry, Zhengzhou 450001, China}

\author{Hui Jing}
\email{jinghui73@gmail.com}
\affiliation{Key Laboratory of Low-Dimensional Quantum Structures and Quantum Control of Ministry of Education, Department of Physics and Synergetic Innovation Center for Quantum Effects and Applications, \\
Hunan Normal University, Changsha 410081, China}

\date{\today}

\begin{abstract}	
The preparation of highly entangled states involving multiparticle systems is of crucial importance in quantum physics, playing a fundamental role in exploring the nature of quantum mechanics and offering essential quantum resources for nascent quantum technologies that surpass classical limits. Here we present how to generate and manipulate tripartite entangled state of photons, phonons, and magnons within a hybrid cavity magnomechanical system. It is shown that by simultaneously applying two coherent driving fields to this system in opposite input directions, it enables a coherent and effective way to regulate the magnomechanical interaction by tuning the phase difference of the driving fields. Based on this feature, it is found that the tripartite entanglement also becomes phase-dependent and can be enhanced for certain phase difference. More interestingly, it is shown that the robustness of tripartite entanglement against environmental thermal noises can also be improved by choosing proper phase difference of the driving fields. Our findings open up a promising way to manipulate and protect fragile tripartite entanglement, which is applicable to a wide range of quantum protocols that require multipartite entangled resources such as quantum communication and quantum metrology.
\end{abstract}

\maketitle

\section{Introduction}\label{Int}
Cavity magnomechanical (CMM) system is a hybrid quantum system composed of optical cavity, magnon and mechanical modes, which provides an excellent platform for studying many interesting physical phenomena~\cite{Tabuchi2014prl,Lachance2017sa,dany2019ape,Wolski2020prl,yuanhypr2022,heqiongyi2023jap,lijie2024njp}. In a hybrid CMM system, the collective spin excitations (i.e., magnons) in the Kittle mode of ferromagnetic crystals, e.g., yttrium-iron-garnet (YIG) spheres, show remarkable experimental compatibility and adjustability~\cite{zouchangling2014prl,Potts2021prx,Rameshti2022pr}. The hybrid CMM system provides a versatile platform and has a wide range of applications, encompassing quantum parametric amplification~\cite{wangyan2023scpma}, magnomechanical chaos~\cite{wuying2019ol,wuying2019lpl,xiong2024pra}, quantum states~\cite{lijie2019njp,lijie2021qst,lijie2023nsr}, mechanical cooling~\cite{Twamley2022prl}, and magnonic frequency comb~\cite{liuzx2022pres,dongchunhua2023prlfc,xionghao2023prafc,xionghao2023frfc}. In particular, CMM-based quantum entanglement~\cite{lijie2018prlcmm,wangyipulijie2023pra,Xiongwei2023prb,xuyejun2024oe,liuwuming2024pra} is a unique quantum resource, has been demonstrated experimentally for various mechanical oscillators~\cite{riedinger2018Remote,OckeloenKorppi2018,kotler2021Direct,mercierdelepinay2021Quantum} and optical fields~\cite{barzanjeh2019Stationarya,chen2020Entanglementa}. However, quantum entanglement typically exhibits weakness and is prone to being disrupted by random noises. Consequently, in practical applications, there is an immediate and pressing necessity to protect and enhance quantum entanglement.

In this paper, we present a scheme to control and enhance tripartite quantum entanglement among the cavity, magnon and mechanical modes within a hybrid quantum CMM system, by applying two pump lasers propagating in different directions. We note that asymmetric optical transmission has been experimentally realized by adjusting the phase difference between the two pump lasers~\cite{2023shilieACS,chen2021Synthetica}. Other quantum effects can also be realized by utilizing dual-drive systems, such as quantum entanglement~\cite{2023Phasecontrolled,Sun2017njp,2024PhasecontrolledPRA}, photon blockade~\cite{Xu2020,Li2022a}, and optical nonreciprocity~\cite{Yan2019ab,Xu2015}. Inspired by these preceding works, here we demonstrate that this scheme can also be used to protect or even enhance tripartite quantum entanglement among the cavity, magnon and mechanical modes in a hybrid quantum CMM system. We find that the tripartite quantum entanglement among the cavity, magnon and mechanical modes in such a hybrid quantum CMM system can be effectively protected and even significantly enhanced, together with improved robustness against environmental thermal noises by properly adjusting the phase difference of the double-pump lasers. Our findings are completely within the current experimental capability and well compatible with other existing methods to protect quantum entanglement~\cite{Xia2023pra,Jiao2020,Gneiting2019,Lai2022,Lii2020pra,Lvv2018pra,Lai2021pra,Qin2018prl,Lvvv2016pra,Xia2020oe,Jiao2024lpr,Lijie2021prresear,2022jiaoPrapp,lipengbo2023prl}, shed a light on fabricating a variety of phase-controlled quantum CMM systems for potential applications in quantum information processing and quantum precision measurements.

This paper is organized as follows. In Sec.\,\ref{SecII}, we introduce the theoretical model of our hybrid CMM system and study its quantum dynamics in detail. In Sec.\,\ref{SecIII}, we explore the phase-controlled tripartite entanglement among the cavity, magnon and mechanical modes and examine its robustness against thermal noises. Finally a brief summary of this article is given in Sec.\,\ref{SecIV}.

\begin{figure*}
	\includegraphics[scale=0.95]{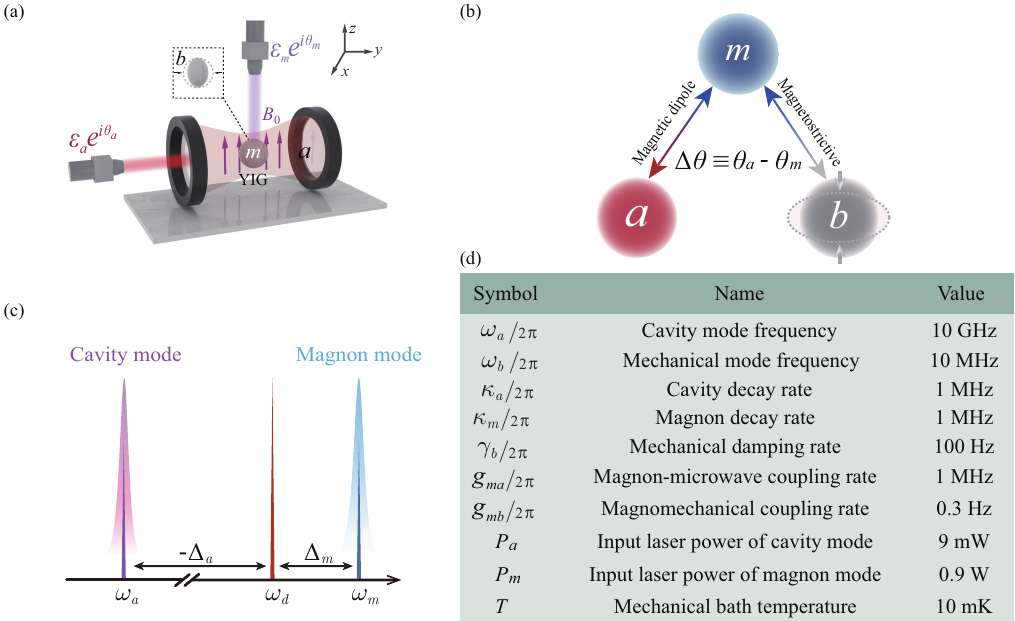}
		\caption{Robust tripartite entanglement among the cavity, magnon and mechanical modes. (a) Schematic diagram of the CMM system with double-pump lasers scheme. A highly polished YIG sphere is placed inside a three-dimensional microwave cavity, near the maximum magnetic field of the cavity, and simultaneously in a uniform static bias magnetic field of strength $B_{0}$ along the $z$-axis, establishing the magnon-photon coupling. (b) Cartoon diagram of the interaction among the cavity, magnon and mechanical modes. (c) Frequency spectrum of this CMM system in panel (a). (d) Experimentally accessible parameters used in the numerical calculations~\cite{you2017nc,tang2016sa}.}\label{Fig1}
\end{figure*}

\section{Phase-controlled CMM system: quantum dynamics}\label{SecII}

In this paper, we investigate how to achieve coherent control and enhancement of the robust tripartite entanglement among the cavity (photon), magnon and mechanical (phonon) modes by adjusting the relative phase of the double-pump lasers. Specifically, as schematically shown in Fig.\,\ref{Fig1}(a), we consider a hybrid CMM system, in which a highly polished YIG sphere ($250\ \textrm{µm}$ diameter) is placed in a three-dimensional microwave cavity~\cite{tang2016sa}. The Kittel mode (magnon mode) of the YIG sphere can be excited via a microwave drive field~\cite{Kittel1948prl}. In addition, a uniform static bias magnetic field of strength $B_{0}$, is applied to the YIG sphere to saturate the magnetization and establish coupling between the magnetic and microwave modes~\cite{HouLiu2019prl}. As shown in Fig.\,\ref{Fig1}(b), the magnon mode couples to the microwave cavity mode via the magnetic dipole interaction. The frequency of magnon mode $\omega_{m}$ is directly determined by the gyromagnetic ratio ($\gamma/2\pi=28\ \textrm{GHz/T}$) and the external bias magnetic field~\cite{yjq2019scpma}, i.e., $\omega_{m}=\gamma$$B_{0}$. Due to the magnetostrictive effect of the YIG sphere~\cite{Bauer2014ssc,hsu1970apl,LeCraw1958prl}, the magnetization changes caused by magnon excitation lead to deformation of the YIG sphere. Meanwhile, the deformation of the YIG sphere in response to an external magnetic field also influences the magnetization, resulting in coupling between the magnon and vibrational (phonon) modes~\cite{tang2016sa,Yamaguchi2022prapp,dchpmp2022prl}. Hence, the YIG sphere supports both the magnon and phonon modes~\cite{tang2016sa}, which are coupled through a nonlinear magnetostrictive interaction [see Fig.\,\ref{Fig1}(b)]. All of these interactions in Fig.\,\ref{Fig1}(b) have been demonstrated experimentally~\cite{wyipu2024prl,dchpmp2022prl,yjqa2022prl,Davis2021prx}. In a frame rotating with driving frequency $\omega_{d}$, the effective Hamiltonian of this CMM system can be written as:
\begin{align}
	\nonumber
	\hat{H}=&~\hat{H}_{0}+\hat{H}_{\mathit{\textrm{int}}}+\hat{H}_{\textrm{dr}},\\ \nonumber
	\hat{H}_{0}=&~\hbar\Delta_{a}\hat{a}^{\dagger}\hat{a}+\hbar\Delta_{m}\hat{m}^{\dagger}\hat{m}+\hbar\frac{\omega_{b}}{2}\left(\hat{q}^{2}+\hat{p}^{2}\right),\\ \nonumber
	\hat{H}_{\textrm{int}}=&~\hbar g_{mb}\hat{m}^{\dagger}\hat{m}\hat{q}+\hbar g_{ma}\left(\hat{a}^{\dagger}\hat{m}+\hat{a}\hat{m}^{\dagger}\right),\\
	\hat{H}_{\textrm{dr}}=&~i\hbar\left(\varepsilon_{a}\hat{a}^{\dagger}e^{-i\theta_{a}}+\varepsilon_{m}\hat{m}^{\dagger}e^{-i\theta_{m}}-\mathrm{H.c.}\right),\label{eq1}
\end{align}
where $\hat{a}$ $(\hat{a}^{\dagger})$ and $\hat{m}$ $(\hat{m}^{\dagger})$ are respectively the annihilation (creation) operators of the cavity mode (with frequency $\omega_{a}$) and magnon mode (with frequency $\omega_{m}$). $\Delta_{a}=\omega_{a}-\omega_{d}$ denotes the optical detuning between the cavity mode and the driving cavity field. $\Delta_{m}=\omega_{m}-\omega_{d}$ denotes the magnon-drive detuning between the magnon mode and the driving magnetic field. $\hat{q}$ and $\hat{p}$ are the dimensionless position and momentum quadratures of the mechanical mode at the resonance frequency $\omega_{b}$. The parameter $g_{ma}$ and $g_{mb}$ denote magnon-microwave coupling rate and single-magnon
magnomechanical coupling rate, respectively. The phase and amplitude of the driving cavity field are given by $\theta_{a}$ and $\left|\varepsilon_{a}\right|=\sqrt{2\kappa_{a}P_{a}/\hbar\omega_{d}}$, where $P_{a}$ is the input laser power of the cavity mode, and $\kappa_{a}$ is the dissipation rate of the cavity mode. The phase and amplitude of the magnetic driving field are characterized by $\theta_{m}$ and $\left|\varepsilon_{m}\right|=\sqrt{2\kappa_{m}P_{m}/\hbar\omega_{d}}$, and $P_{m}$ is referred to as the input laser power of the magnon mode, and $\kappa_{m}$ denotes the dissipation rate of the magnon mode~\cite{dujiangfneg2021prapp}. The phase difference of the double-pump lasers is defined through $\Delta\theta\equiv\theta _{a}-\theta_{m}$. By adjusting this phase difference $\Delta\theta$, the robust tripartite entanglement among the cavity, magnon, and mechanical modes can be well controlled.

By considering the influence of this CMM system dissipations and environmental input noises, the dynamic evolutions of the system can be fully characterized by the quantum Langevin equations (QLEs) as:
\begin{align}
	\nonumber
	\dot{\hat{a}}=&-\left(i\Delta_{a}+\kappa_{a}\right)\hat{a}-ig_{ma}\hat{m}+\varepsilon_{a}e^{-i\theta_{a}}+\sqrt{2\kappa_{a}}\hat{a}^{in},\\ \nonumber
	\dot{\hat{m}}=&-\left(i\Delta_{m}+\kappa_{m}\right)\hat{m}-ig_{ma}\hat{a}-ig_{mb}\hat{m}\hat{q}+\varepsilon_{m}e^{-i\theta_{m}}\\ \nonumber
	&+\sqrt{2\kappa_{m}}\hat{m}^{in},\\ \nonumber
	\dot{\hat{q}}=&~\omega_{b}\hat{p},\\
	\dot{\hat{p}}=&-\hat{q}\omega_{b}-g_{mb}\hat{m}^{\dagger}\hat{m}-\gamma_{b}\hat{p}+\hat{\xi},
	\label{eq2}
\end{align}
where $\gamma_{b}$ is referred to as the mechanical damping rate. $\hat{a}^{in}$, $\hat{m}^{in}$, and $\hat{\xi}$ denote the input vacuum noise operators for the cavity, magnon, and mechanical modes, respectively. These input vacuum noise operators have zero-mean values, which are characterized by the following correlation functions~\cite{Gardiner2000}:
\begin{align}
	\nonumber
	\left\langle \hat{a}^{in}(t)\hat{a}^{in\dagger}\left(t^{\prime}\right)\right\rangle&=\left[N_{a}\left(\omega_{a}\right)+1\right]\delta\left(t-t^{\prime}\right),\\  
	\left\langle \hat{a}^{in\dagger}(t)\hat{a}^{in}\left(t^{\prime}\right)\right\rangle&=N_{a}\left(\omega_{a}\right)\delta\left(t-t^{\prime}\right),\\ \nonumber
	\left\langle \hat{m}^{in}(t)\hat{m}^{in\dagger}\left(t^{\prime}\right)\right\rangle&=\left[N_{m}\left(\omega_{m}\right)+1\right]\delta\left(t-t^{\prime}\right),\\ \nonumber
	\left\langle \hat{m}^{in\dagger}\left(t\right)\hat{m}^{in}(t^{\prime})\right\rangle&=N_{m}\left(\omega_{m}\right)\delta\left(t-t^{\prime}\right),\\ 
	\left\langle \xi(t)\xi\left(t^{\prime}\right)+\xi\left(t^{\prime}\right)\xi(t)\right\rangle /2&\simeq\gamma_{b}\left[2N_{b}\left(\omega_{b}\right)+1\right]\delta\left(t-t^{\prime}\right),\nonumber 
\end{align}
where $N_{j}(\omega_{m_{j}})=[\exp(\hbar\omega_{m_{j}}/k_{\textit{B}}T)-1]^{-1} (j=a, m, b)$ denotes respectively the mean thermal excitation number of the photon, magnon, and phonon, with $k_{\textit{B}}$ the Boltzmann constant and $T$ the bath temperature.

Under the conditions of strong driving fields, each operator for this CMM system can be expanded into the form of the sum of its steady-state mean and a small quantum fluctuation around it, i.e.,
\begin{align}
	\nonumber
	\hat{a}&=\alpha_{s}+\delta\hat{a},~~~~\hat{m}=m_{s}+\delta\hat{m}\\
	\hat{q}&=q_{s}+\delta\hat{q},~~~~~~\hat{p}=p_{s}+\delta\hat{p}.
	\label{eq4}
\end{align}
By substituting Eq.\,(\ref{eq4}) into QLEs\,(\ref{eq2}), we can obtain the first-order inhomogeneous differential equations for steady-state mean values, i.e.,
\begin{align}
	\nonumber
	\dot{\alpha}_{s}=&-\left(i\Delta_{a}+\kappa_{a}\right)\alpha_{s}-ig_{ma}m_{s}+\varepsilon_{a}e^{-i\theta_{a}},\\ \nonumber
	\dot{m}_{s}=&-\left(i\Delta_{m}+\kappa_{m}\right)m_{s}-ig_{ma}\alpha_{s}-ig_{mb}m_{s}q_{s}\\ \nonumber
	&+\varepsilon_{m}e^{-i\theta_{m}}\\ \nonumber
	\dot{q}_{s}=&~\omega_{b}p_{s},\\
	\dot{p}_{s}=&-q_{s}\omega_{b}-g_{mb}\left|m_{s}\right|^2.
	\label{eq5}
\end{align}
The corresponding linearized QLEs for quantum fluctuations can be written as
\begin{align}
	\nonumber
	\delta\dot{\hat{a}}=&-\left(i\Delta_{a}+\kappa_{a}\right)\delta\hat{a}-ig_{ma}\delta\hat{m}+\sqrt{2\kappa_{a}}\hat{a}^{in},\\ \nonumber
	\delta\dot{\hat{m}}=&-\left(i\Delta_{m}+\kappa_{m}\right)\delta\hat{m}-ig_{ma}\delta\hat{a}-ig_{mb}m_{s}\delta\hat{q}\\ \nonumber
	&-ig_{mb}\delta\hat{m}{q}_{s}+\sqrt{2\kappa_{m}}\hat{m}^{in},\\ \nonumber
	\delta\dot{\hat{q}}=&~\omega_{b}\delta\hat{p},\\
	\delta\dot{\hat{p}}=&-\omega_{b}\delta\hat{q}-g_{mb}m_{s}^{\dagger}\delta\hat{m}-g_{mb}m_{s}\delta\hat{m}^{\dagger}-\gamma_{b}\delta\hat{p}+\hat{\xi}.
	\label{eq6}
\end{align}
By setting all the derivatives in Eq.\,(\ref{eq5}) as zero, the steady-state mean values of each mode can be obtained 
\begin{align}
	\nonumber
	\alpha_s=&~-\frac{ig_{ma}m_{s}-\varepsilon_a e^{-i \theta_a}}{i \Delta_{a}+\kappa_{a}}, \\ \nonumber
	m_{s}=&~\frac{-i g_{ma} \varepsilon_a e^{-i \theta_a}+ \left(i \Delta_{a} +\kappa_{a} \right)\varepsilon_m e^{-i \theta_m}}{\left(i \tilde\Delta_{m}+\kappa_{m}\right)\left(i \Delta_{a}+\kappa_{a}\right)+g_{ma}^2},\\
	q_{s}=&~-\frac{g_{mb}\left|m_{s}\right|^2}{\omega_{b}},
	\label{eq7}
\end{align}
where $\tilde\Delta_{m}=\Delta_{m}+g_{mb}q_{s}$ denotes the effective magnon-drive detuning. Furthermore, due to $|\tilde\Delta_{m}|$, $|\Delta_{a}|$ $\!\gg\!$ $\kappa_{a}$, $\kappa_{m}$, the steady-state mean values of the magnon mode $m_{s}$ can be written as the following simple form:
\begin{align}
	m_{s}=\frac{-i g_{ma} \varepsilon_a e^{-i \theta_a}+i\Delta_{a}\varepsilon_m e^{-i \theta_m}}{g_{ma}^2-\tilde\Delta_{m}\Delta_{a}}.
	\label{eq8}
\end{align}
Defining the optical and magnon quadrature fluctuations operators as
\begin{align}
	\nonumber
	\delta \hat{X}&=\frac{\delta \hat{a}+\delta \hat{a}^{\dagger}}{\sqrt{2}},~~~~~~~~~~~~\delta \hat{Y}=\frac{\delta \hat{a}-\delta \hat{a}^{\dagger}}{i\sqrt{2}},\\
	\delta \hat{x}&=\frac{\delta \hat{m}+\delta \hat{m}^{\dagger}}{\sqrt{2}},~~~~~~~~~~\delta \hat{y}=\frac{\delta \hat{m}-\delta \hat{m}^{\dagger}}{i\sqrt{2}},
	\label{eq9}
\end{align}
and the associated input noise operators as
\begin{align}
	\nonumber
	\delta \hat{X}^{in}&=\frac{\delta \hat{a}^{in}+\delta \hat{a}^{in,\dagger}}{\sqrt{2}},~~~~\delta \hat{Y}^{in}=\frac{\delta \hat{a}^{in}-\delta \hat{a}^{in,\dagger}}{i\sqrt{2}},\\
	\delta \hat{x}^{in}&=\frac{\delta \hat{m}^{in}+\delta \hat{m}^{in,\dagger}}{\sqrt{2}},~~\delta \hat{y}^{in}=\frac{\delta \hat{m}^{in}-\delta \hat{m}^{in,\dagger}}{i\sqrt{2}},
	\label{eq10}
\end{align}
the corresponding linearized QLEs can be written explicitly in a compact form as
\begin{align}
	\dot{\hat{u}}(t)=A\hat{u}(t)+\hat{v}(t),\label{eq11}
\end{align}
where we have introduce the fluctuation operator vector $\hat{u}$
\begin{align}
	\hat{u}^T(t)=(\delta \hat{X}, \delta \hat{Y}, \delta \hat{x}, \delta \hat{y}, \delta \hat{q}, \delta \hat{p}),
	\label{eq12}
\end{align}
the input noise operator $\hat{v}$
\begin{align}
	\nonumber
	\hat{v}^T(t)=(&\sqrt{2\kappa_{a}}\hat{X}^{\textrm{in}}, \sqrt{2\kappa_{a}}\hat{Y}^{\textrm{in}}, \sqrt{2\kappa_{m}}\hat{x}^{\textrm{in}}, \\ 
	&\sqrt{2\kappa_{m}}\hat{y}^{\textrm{in}}, 0, \hat{\xi}),
	\label{eq13}
\end{align}
and the corresponding coefficient matrix $A$ is given by
\begin{align}
	A=\begin{pmatrix}
		-\kappa_a & \Delta_a & 0 & g_{ma} & 0 & 0\\
		-\Delta_a & -\kappa_a & -g_{ma} & 0 & 0 & 0\\
		0 & g_{ma} & -\kappa_m & \tilde{\Delta}_m & -G_{mb} & 0\\
		-g_{ma} & 0 & -\tilde{\Delta}_m & -\kappa_m & 0 & 0\\
		0 & 0 & 0 & 0 & 0 & \omega_{m}\\
		0 & 0 & 0 & G_{mb} & -\omega_{b} & -\gamma_{b}
	\end{pmatrix},
	\label{eq14}
\end{align}
with the component $G_{mb}=i\sqrt{2}g_{mb}m_{s}$ denotes the effective magnomechanical coupling rate. The effective magnomechanical coupling rate can be effectively adjusted by the phase difference (i.e., $\Delta\theta\equiv\theta _{a}-\theta_{m}$), which forms the basis for the manipulation of the tripartite entanglement among photons, phonons, and magnons. Also, the solution of the linearized QLEs (\ref{eq11}) is given by
\begin{align}
	\hat{u}(t)=\mathcal{M}(t)\hat{u}(0)+\int_{0}^{t}d\tau \mathcal{M}(\tau)\hat{v}(t-\tau),\label{eq15}
\end{align}
where
\begin{align}
	\mathcal{M}(t)=\exp(At).\label{eq16}
\end{align}
The system is stable only when all real part of the eigenvalues of $A$ is negative, as characterized by the Routh-Hurwitz criterion~\cite{dejesus1987RouthHurwitza}. When all the stability conditions are fulfilled, we can obtain $\mathcal{M}(\infty)=0$ in the steady state, and
\begin{align}
	\hat{u}_{i}(\infty)=\int_{0}^{\infty}d\tau\sum_{k}\mathcal{M}_{ik}(\tau)\hat{v}_{k}(t-\tau).\label{eq17}
\end{align}
Because of the linearized dynamics and the Gaussian nature of the quantum noise, the steady state of the quantum fluctuations of in this CMM system can finally evolve into a quadripartite zero-mean Gaussian state, which is fully characterized by a $6\times6$ correlation matrix (CM) $V$ with the components
\begin{align}
	V_{kl}=\langle \hat{u}_{k}(\infty)\hat{u}_{l}(\infty)\!+\!\hat{u}_{l}(\infty)\hat{u}_{k}(\infty)\rangle/2. \label{eq18}
\end{align}
By substituting Eq.\,(\ref{eq17}) into Eq.\,(\ref{eq18}) and using the fact that the six components of $\hat{v}(t)$ are not correlated with each other, the steady-state CM $ V $ is obtained by
\begin{align}
	V=\int_{0}^{\infty}d\tau\mathcal{M}(\tau)D\mathcal{M}^{\text{T}}(\tau), \label{19}
\end{align}
where
\begin{align}
\nonumber
    D\!=~\!\textrm{Diag}\,[&\kappa_a(2N_{a}+1),\kappa_a(2N_{a}+1),\kappa_m(2N_{m}+1),\\
    &\kappa_m(2N_{m}+1),0,\gamma_{b}(2N_{b}+1)], \label{eq20}
\end{align}
is a diffusion matrix, which is obtained through
\begin{align}
	\langle \hat{v}_{k}(\tau)\hat{v}_{l}(\tau')\!+\!\hat{v}_{l}(\tau')\hat{v}_{k}(\tau)\rangle/2=D_{kl}\delta(\tau-\tau').\label{eq21}
\end{align}
Under the stability condition, the steady-state CM $V$ fulfills the Lyapunov equation~\cite{vitali2007Optomechanical}:
\begin{align}
	AV+VA^{\text{T}}=-D. \label{eq22}
\end{align}
The Lyapunov equation (\ref{eq22}) is a linear equation and allows us to find CM $V$ for any values of the relevant parameters. However, the analytical expression of $V$ is too complicated and thus is not reported here.

\begin{figure}
	\centering
	\includegraphics[scale=0.95]{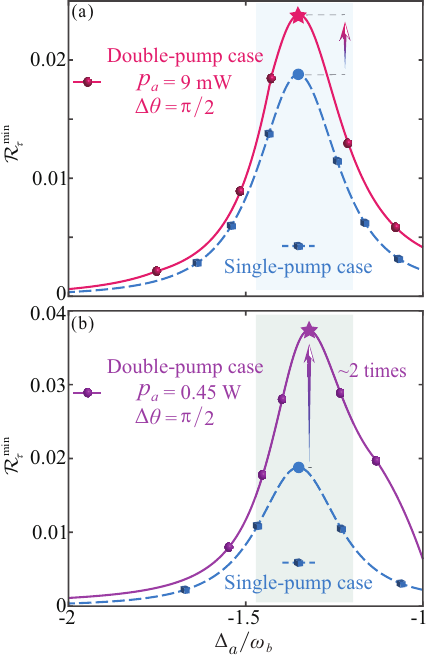}
	\caption{The degree of tripartite CMM entanglement could be significantly enhanced for the double-pump case in comparison to the single-pump case. (a) The minimum residual contangle $\mathcal{R}^{\textrm{min}}_{\tau}$ versus the optical detuning $\Delta_{a}/\omega_{b}$, with $P_{a}=9~\mathrm{mW}$ and the phase difference $\Delta\theta=\pi/2$ of driving fields. Compared with the single-pump case, the maximum value of $\mathcal{R}^{\textrm{min}}_{\tau}$ could be significantly enhanced for the double-pump case. (b) The minimum residual contangle $\mathcal{R}^{\textrm{min}}_{\tau}$ versus the optical detuning $\Delta_{a}/\omega_{b}$, with $P_{a}=0.45~\mathrm{W}$ and the phase difference $\Delta\theta=\pi/2$. Compared with the single-pump case, the maximum value of $\mathcal{R}^{\textrm{min}}_{\tau}$ could be enhanced for $\sim2$ times for the double-pump case. In (a-b), we select $\tilde\Delta_{m}/\omega_{b}=0.9$. The other parameters are chosen as the same in the table of Fig.\,\ref{Fig1}(d).} 
	\label{Fig2}
\end{figure}
\begin{figure*}
	\centering
	\includegraphics[scale=0.95]{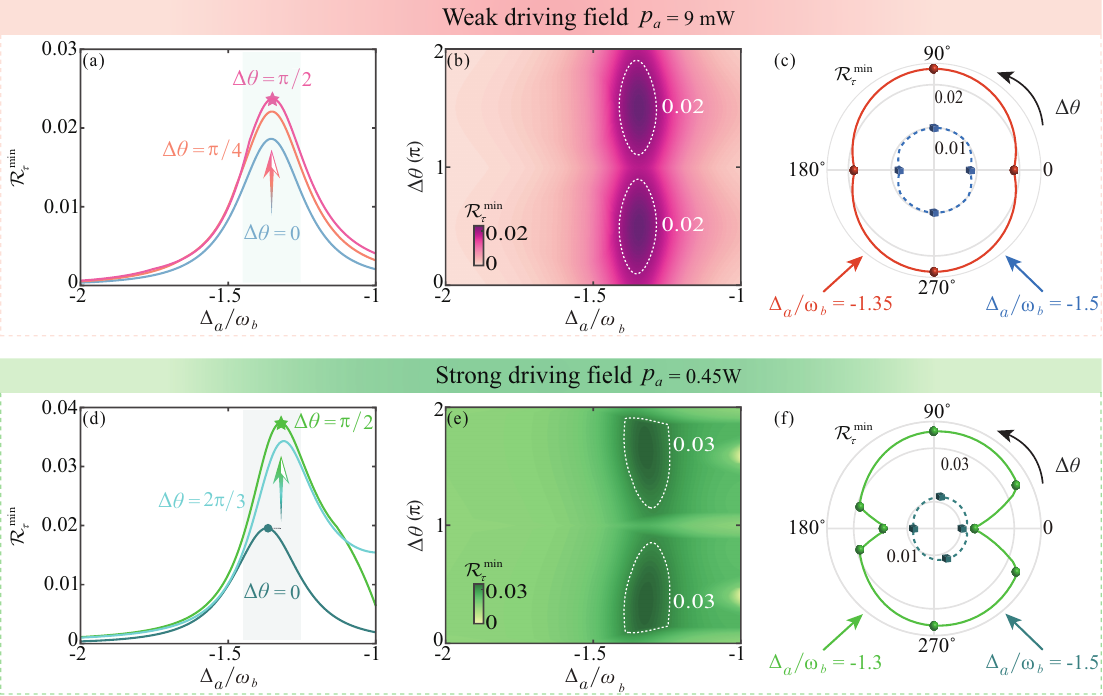}
	\caption{Phase-controlled robust tripartite entanglement among the cavity, magnon and mechanical modes by tuning phase difference $\Delta\theta$ of the driving lasers. (a), (d) The minimum residual contangle $\mathcal{R}^{\textrm{min}}_{\tau}$ versus the optical detuning $\Delta_{a}/\omega_{b}$ for different values of phase difference $\Delta\theta$. (b), (e) Density plot of the minimum residual contangle $\mathcal{R}^{\textrm{min}}_{\tau}$ as a function of the optical detuning $\Delta_{a}/\omega_{b}$ and phase difference $\Delta\theta$. (c), (f) The minimum residual contangle $\mathcal{R}^{\textrm{min}}_{\tau}$ is plotted as a function of phase difference $\Delta\theta$ in polar coordinates, (c) with $\Delta_{a}/\omega_{b}=-1.35$ or $-1.5$, and (f) $\Delta_{a}/\omega_{b}=-1.3$ or $-1.5$. (a-c) We adopt weak driving field $P_{a}=9~\mathrm{mW}$. (d-f) We adopt strong driving field $P_{a}=0.45~\mathrm{W}$. The other parameters are chosen as the same in Fig.\,\ref{Fig2}.} 
	\label{Fig3}
\end{figure*}

\section{Results and discussions}\label{SecIII}

To study the tripartite quantum entanglement of the three-mode CMM system, we apply a quantitative measure of the minimum residual contangle $\mathcal{R}^{\textrm{min}}_{\tau}$~\cite{Illuminati2006njp,Illuminati2007jpa}, which is defined as
\begin{align}
	\mathcal{R}^{\textrm{min}}_{\tau}\equiv\min_{(r,s,t)}\,\big[E^{r|st}_{\tau}-E^{r|s}_{\tau}-E^{r|t}_{\tau}\big],
	\label{eq23}
\end{align}
where $(r,s,t)\in\{a,m,b\}$ denotes all the possible permutations of the three-mode indexes. $E_{\tau}^{u|v}$ is the contangle of subsystems of $u$ ($u$ contains one mode) and $v$ ($v$ contains one or two modes), which can be defined by a proper entanglement monotone, e.g., the squared logarithmic negativity.

\begin{align}
	E^{u|v}_{\tau}\equiv\big[E_{N}\big]^2\equiv\left\{\max\,[0,-\ln (2\tilde\nu_{-})]\right\}^2,
	\label{eq24}
\end{align}
Here, $\tilde\nu_{-}$ denotes the minimum symplectic eigenvalue of the CM.

If $v$ contains only one mode, $\tilde\nu_{-}$ in Eq.\,(\ref{eq24}) is given by
\begin{align}
	\tilde\nu_{-}=\min\,\big[\textrm{eig}|i\Omega_{2}\tilde{V}_{4}|\big],
	\label{eq25}
\end{align}
where
\begin{align}
	\Omega_{2}=\oplus_{j=1}^{2}i\sigma_{y},
	\label{eq26}
\end{align}
where $\sigma_{y}$ denotes the y-Pauli matrix. The matrix $\tilde{V}_{4}$ in Eq.\,(\ref{eq25}) is given by
\begin{align}
	\tilde{V}_{4}=P_{0}V_{4}P_{0},
	\label{eq27}
\end{align}
where $V_{4}$ denotes the $4\times4$ CM of two subsystems, obtained by removing the rows and columns of the uninteresting mode in $V$, and  
\begin{align}
	P_{0}=\textrm{diag}(1,-1,1,1),
	\label{eq28}
\end{align}
is the matrix that realizes the partial transposition at the level of the CMs.

If $v$ contains two modes, $\tilde\nu_{-}$ in Eq.\,(\ref{eq24}) is obtained by

\begin{align}
	\tilde\nu_{-}=\min\,\big[\textrm{eig}|i\Omega_{3}\tilde{V}_{6}|\big],
	\label{eq29}
\end{align}
where
\begin{align}
	\Omega_{3}=\oplus_{j=1}^{3}i\sigma_{y},
	\label{eq30}
\end{align}
and, the matrix $\tilde{V}_{6}$ in Eq.\,(\ref{eq29}) is given by
\begin{align}
	\tilde{V}_{6}=P_{0}V_{6}P_{0},
	\label{eq31}
\end{align}
where $V_{6}$ denotes the $6\times6$ CM of this system, and we introduce the partial transposition matrices
\begin{align}
	\nonumber
	P_{a|mb}=\textrm{diag}(1,-1,1,1,1,1),\\ \nonumber
	P_{m|ab}=\textrm{diag}(1,1,1,-1,1,1),\\
	P_{b|am}=\textrm{diag}(1,1,1,1,1,-1).
	\label{eq32}
\end{align}
The residual contangle satisfies the monogamy of quantum entanglement, i.e.,
\begin{align}
	E^{r|st}_{\tau}-E^{r|s}_{\tau}-E^{r|t}_{\tau}\geq0.
	\label{eq33}
\end{align}
This inequality is similar to the Coffman-Kundu-Wootters monogamy inequality~\cite{Wootters2000pra}, which holds for three qubits. The nonzero minimum residual contangle $\mathcal{R}_{\tau}^{\textrm{min}}>0$ means that the full tripartite inseparability is generated.

Now we study the tripartite entanglement of this hybrid CMM system in detail. In Fig.\,\ref{Fig2}, we first demonstrate that the degree of tripartite CMM entanglement could be significantly enhanced for the double-pump case in comparison to the single-pump case. In our numerical calculations, to ensure the stability and experimental feasibility of this CMM system, the following parameters are employed~\cite{you2017nc,tang2016sa}: $\omega_{a}/2\pi=10\,\textrm{GHz}$, $\omega_{b}/2\pi=10\,\textrm{MHz}$, $\kappa_{a}/2\pi=1\,\textrm{MHz}$, $\kappa_{m}/2\pi=1\,\textrm{MHz}$, $\gamma_{b}/2\pi=100\,\textrm{Hz}$, $g_{ma}/2\pi=1\,\textrm{MHz}$, $g_{mb}/2\pi=0.3\,\textrm{MHz}$, $\lambda=780\,\textrm{nm}$, $P_a=9\,\textrm{mW}$, $P_m=0.9\,\textrm{W}$, and $T=10\,\textrm{mK}$. Specifically, as shown in Fig.\,\ref{Fig2}, the minimum residual contangle $\mathcal{R}^{\textrm{min}}_{\tau}$ is plotted as a function of the optical detuning $\Delta_{a}/\omega_{b}$ for different values of the input laser power $P_{a}$ of the cavity mode in the double-pump laser case. We also plot the minimum residual contangle $\mathcal{R}^{\textrm{min}}_{\tau}$ as a function of the optical detuning $\Delta_{a}/\omega_{b}$ for the single-pump laser case [see Fig.\,\ref{Fig2}]. As shown in Fig.\,\ref{Fig2}, for a special value of the phase difference $\Delta\theta$ of the two driving lasers, e.g., $\Delta\theta=\pi/2$, compared with single-pump, double-pump can generate more robust tripartite CMM entanglement. In particular, under the strong driving field $P_{a}=0.45~\mathrm{W}$ condition, the minimum residual contangle $\mathcal{R}^{\textrm{min}}_{\tau}$ could be enhanced for $\sim2$ times in comparison with that of a single driving laser [see Fig.\,\ref{Fig2}(b)].

Furthermore, in Fig.\,\ref{Fig3}, we explore how to regulate the behavior of the tripartite CMM entanglement among the cavity, magnon and mechanical modes by adjusting the phase difference $\Delta\theta$ of the two driving lasers. Meanwhile, we further study the influence of the phase difference $\Delta\theta$ on the generation of tripartite CMM entanglement for different input laser power $P_{a}$ of the cavity mode. Specifically, as shown in Figs.\,\ref{Fig3}(a) and \ref{Fig3}(d), we plot the minimum residual contangle $\mathcal{R}^{\textrm{min}}_{\tau}$ as a function of the optical detuning $\Delta_{a}/\omega_{b}$ for different values of the phase difference $\Delta\theta$. It is seen that the degree of such tripartite CMM entanglement is the greatest near the optical detuning $\Delta_{a}/\omega_{b}\simeq1.3$. The minimum residual contangle $\mathcal{R}^{\textrm{min}}_{\tau}$ can be effectively modulated or even enhanced by adjusting the phase difference $\Delta\theta$ of the driving lasers [see Figs.\,\ref{Fig3}(a) and \ref{Fig3}(d)]. In particular, in the case of strong driving field $P_{a}=0.45~\mathrm{W}$, the degree of tripartite CMM entanglement could be considerably enhanced for the phase difference $\Delta\theta=\pi/2$ case in comparison with that of the case phase difference $\Delta\theta=0$ [see Fig.\,\ref{Fig3}(d)]. In order to see the regulation effect of the phase difference $\Delta\theta$ on tripartite CMM entanglement more clearly, we also show the dependence of the minimum residual contangle $\mathcal{R}^{\textrm{min}}_{\tau}$ on the optical detuning $\Delta_{a}/\omega_{m}$ and the phase difference $\Delta\theta$ in Figs.\,\ref{Fig3}(b) and \ref{Fig3}(e). It is seen that $\mathcal{R}^{\textrm{min}}_{\tau}$ is well regulated periodically by tuning the phase difference $\Delta\theta$ of the driving lasers [see Figs.\,\ref{Fig3}(b-c) and \ref{Fig3}(e-f)].

\begin{figure}
	\centering
	\includegraphics[scale=0.95]{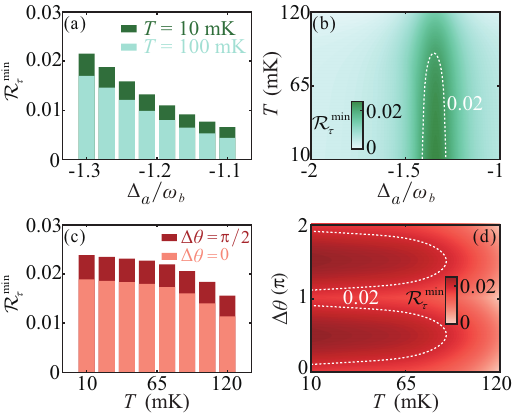}
	\caption{The influence of thermal effects on tripartite CMM entanglement. (a) The minimum residual contangle $\mathcal{R}^{\textrm{min}}_{\tau}$ versus the optical detuning $\Delta_{a}/\omega_{b}$ for different environment temperature $T$. (b) Density plot of the minimum residual contangle $\mathcal{R}^{\textrm{min}}_{\tau}$ as a function of  the optical detuning $\Delta_{a}/\omega_{b}$ and the environment temperature $T$. Here, we have chosen the phase difference $\Delta\theta=\pi/2$. (c) The minimum residual contangle $\mathcal{R}^{\textrm{min}}_{\tau}$ versus the environment temperature $T$ for different phase difference $\Delta\theta$ of driving fields. (d) Density plot of the minimum residual contangle $\mathcal{R}^{\textrm{min}}_{\tau}$ as a function of the environment temperature $T$ and the phase difference $\Delta\theta$. Here, we have chosen the optical detuning $\Delta_{a}/\omega_{b}=1.35$. The other parameters are chosen as the same in Fig.\,\ref{Fig3}(a).} 
	\label{Fig4}
\end{figure}

Finally, we also confirm the influence of the phase difference $\Delta\theta$ on tripartite CMM entanglement generation for different values of bath temperature $T$. For this purpose, we plot the minimum residual contangle $\mathcal{R}^{\textrm{min}}_{\tau}$ as a function of the optical detuning $\Delta_{a}/\omega_{b}$ in Fig.\,\ref{Fig4}(a), with bath temperature $T=10$ or $100$ mK. Figure\,\ref{Fig4}(b) shows the dependence of the minimum residual contangle $\mathcal{R}^{\textrm{min}}_{\tau}$ on the optical detuning $\Delta_{a}/\omega_{b}$ and the bath temperature $T$, with $\Delta\theta=\pi/2$. We also plot the minimum residual contangle $\mathcal{R}^{\textrm{min}}_{\tau}$ as a function of the bath temperature $T$ in Fig.\,\ref{Fig4}(c), with $\Delta\theta=\pi/2$ or $0$. Figure\,\ref{Fig4}(d) shows the dependence of the minimum residual contangle $\mathcal{R}^{\textrm{min}}_{\tau}$ on bath temperature $T$ and the phase difference $\Delta\theta$, with $\Delta_{a}/\omega_{m}=1.35$. It is seen that, the tripartite CMM entanglement is fragile to thermal noise, and it always tends to be destroyed when the bath temperature increases. However, it is also seen that, under the same condition of the bath temperature $T$, the tripartite CMM entanglement could become more robust to thermal noise with respect to some specific values of phase difference $\Delta\theta$ of the driving fields. Therefore, our proposed scheme is expected to be used to protect the fragile tripartite entanglement from thermal noise.

\section{Conclusion}\label{SecIV}

In summary, we have studied how to generate, manipulate, and even enhance the tripartite quantum entanglement among the cavity, magnon and mechanical modes by properly adjusting the phase difference $\Delta\theta$ of the double-pump lasers in a hybrid CMM system. We find that the degree of tripartite CMM entanglement can be significantly enhanced, and the robustness of such entanglement against thermal noises can also be improved by tuning the phase difference $\Delta\theta$ of the driving lasers. Our findings, shedding a light on strategies to protect and enhance the performance of various quantum devices in practical noisy environments, provide a compelling opportunity to bring to fruition a range of entanglement-enabled quantum technologies, including quantum networking~\cite{Hermans2022,Simon2017,Komar2014,Shen2009pra}, quantum sensing~\cite{Degen2017,Gilmore2021,Barzanjeh2020,Chen2027nature}, and quantum computing~\cite{Gyongyosi2019,Knill2005,OBrien2007}. In a broader view, we envision that our work can be extended to investigate various other quantum effects based on such CMM systems, such as ultrahigh-sensitive sensing~\cite{Colombano2020Prl,Potts2020prapp,zhangqian2024scpma}, magnon blockade~\cite{zhouling2020prampb,wyp202scpma,whf2024prampb}, ultra-slow light~\cite{huai2019prampb,lu2021prampb,lu2023prampb}.  

\section{Acknowledgments}\label{V}
H. J. is supported by the National Natural Science Foundation of China (NSFC, Grants No. 11935006, and No. 12421005), the National Key R$\&$D Program of China (Grants No. 2024YFE0102400), the Hunan Provincial Major Sci-Tech Program (Grant No. 2023ZJ1010), and the Science and Technology Innovation Program of Hunan Province (Grant No. 2020RC4047). Y.-F. J. is supported by the NSFC (Grant No. 12405029).

\end{document}